\pdfoutput=1
\documentclass[11pt]{article}
\usepackage[margin=1in]{geometry}
\usepackage[T1]{fontenc}
\usepackage{mathptmx}
\usepackage{microtype}
\usepackage{graphicx}
\usepackage{array}
\usepackage{booktabs}
\usepackage[font=small,labelfont=bf,labelsep=period,margin=0.4cm]{caption}
\usepackage[round,authoryear]{natbib}
\setcitestyle{aysep={}}
\usepackage{xcolor}
\definecolor{linkblue}{rgb}{0.10,0.27,0.55}
\usepackage[colorlinks=true,linkcolor=linkblue,citecolor=linkblue,urlcolor=linkblue]{hyperref}
\usepackage{url}
\usepackage{titlesec}
\titleformat{\section}{\large\bfseries}{\thesection}{0.6em}{}
\titleformat{\subsection}{\normalsize\bfseries}{\thesubsection}{0.6em}{}
\titlespacing*{\section}{0pt}{1.4ex plus .4ex}{0.8ex}
\titlespacing*{\subsection}{0pt}{1.1ex plus .3ex}{0.5ex}

\usepackage{fancyhdr}
\title{\Large\bfseries Missed Detections and False Alarms in a CNN Technosignature Detector on Breakthrough Listen Data}
\author{Omer Imran\\[2pt] \small\itshape Independent researcher\\ \small\href{mailto:imomerimran@gmail.com}{imomerimran@gmail.com}}
\date{}

\begin{document}
\maketitle\thispagestyle{plain}

\begin{abstract}
Radio technosignature searches rely on machine learning detectors to shrink the candidate pool before a human looks at it. They are validated on held-out data from the training distribution, which excludes the cases where they are most likely to fail: a technosignature has unknown morphology, and terrestrial interference is unbounded in variety.

I trained a standard competition detector once, froze it, and probed it with 15,000 injections across ten morphologies and five amplitudes, placed into real Green Bank backgrounds. Two directions were tested: genuine signals with morphologies other than a straight narrowband line, and interference resembling the training data.

Recall did not follow the predicted ordering, and it fell as signals brightened, even for a needle of the kind the detector was trained on (from 0.590 to 0.090). One interference morphology reached a false positive rate of 0.627, comparable to the detector's own baseline recall of 0.5525 on genuine signals. Gradient-weighted Class Activation Mapping (Grad-CAM) showed attention along the injected path in one missed family by a pre-specified test, and visually in a second. A pre-specified, time-averaged ON-OFF filter removed false alarms and genuine detections at the same rate.

The detector reads only the ON panels. Since the only evidence distinguishing a genuine candidate from terrestrial interference lies in the discarded OFF panels, it cannot tell if a signal originates from the target star; it can only tell if it resembles the training data. This explains the false alarm directly; the missed signals reflect what the detector learned to recognise, which reading the OFF panels would not by itself change.
\end{abstract}

\section{Introduction}
Radio SETI (Search for Extraterrestrial Intelligence) surveys cover thousands of nearby stars, and each observation produces more candidate signals than a human team can inspect directly. Machine learning detectors are therefore used to reduce the candidate volume before human inspection. A survey of 692 stars \citep{enriquez2017} produced 11 candidates, all of which were consistent with RFI (Radio Frequency Interference) on closer analysis.

These detectors are validated against held-out data from the same distribution they were trained on. The validation relies on an assumption that does not hold in this domain: that what a detector encounters in deployment resembles what it was trained on. A technosignature has an unknown morphology; if its appearance were known, it would not need to be searched for. The same gap applies to interference as RFI is unbounded in variety, so new morphologies constantly appear. The cases on which such a detector is most likely to fail are absent from the cases used to validate it.

This paper measures how such a detector behaves outside that distribution. I trained a detector once using a public baseline from the SETI Breakthrough Listen Kaggle competition \citep{bl2021}, froze all parameters, and probed it in two directions: genuine signals with morphologies other than a straight narrowband line, and interference with morphologies resembling what it was trained on. A total of 15,000 injections were placed into real Green Bank backgrounds across ten morphologies and five amplitudes.

The detector's recall did not follow the predicted ordering, as a morphology predicted to sit mid-way from the training distribution scored the highest, above the two predicted to be closest. One interference morphology produced a false positive rate of 0.627, comparable to its own baseline recall of 0.5525 on genuine signals. Grad-CAM showed attention along the injected path in one missed family by a pre-specified test, and visually in a second. A pre-specified, time-averaged ON-OFF filter removed false alarms and genuine detections at the same rate.

One property of the input accounts for the false alarms. The detector reads only the ON panels and the only evidence distinguishing a genuine candidate from terrestrial interference is the OFF panels, which are discarded before the model even sees the input. Hence it cannot be asking whether a signal originates from the target star; it can only ask whether it resembles its training set. The missed signals have a different source: they reflect what the detector learned to recognise, which a detector reading all six panels could share. That the detector cannot use the OFF panels follows from its input design; what this paper measures is the cost.

\section{Related Work}
Radio technosignature searches have historically relied on de-Doppler search pipelines such as turboSETI \citep{enriquez2017}, which scan for narrowband signals that slowly drift in frequency and retain hits above a signal-to-noise threshold. Candidates are then tested against the ON/OFF cadence, and a genuine signal should appear in the ON observations and be absent from the OFF observations. By contrast, terrestrial interference is expected to appear in both. Surveys of 692 and 1,327 nearby stars \citep{enriquez2017,price2020} used this method and reported upper limits rather than detections. In both cases, every candidate resolved to RFI. Because this approach is defined in terms of narrow drifting lines, any signal of another shape is invisible to it; not missed, but outside what the algorithm can represent. \citet{price2020} noted that their analysis is confined to narrow drifting signals, and named pulsed signals as future work.

The volume of candidates led to machine learning being introduced: the 1,327-star survey produced 51 million hits above threshold, which its classical ON/OFF pipeline narrowed to 6,154 event groups, none of which survived manual inspection. Subsequent work has used several approaches:

\citet{ma2023} applied a convolutional variational autoencoder to 480 hours of observations that covered 820 stars. \citet{harp2019} classified spectrograms as images using a convolutional neural network (CNN). \citet{pinchuk2022} trained a CNN to identify RFI using its direction of origin.

These systems remain built around narrow drifting signals, though not all of them discard the OFF observations: \citet{ma2023} build the ON/OFF cadence into what their pipeline learns. The ON-only design examined in this paper is that of the public competition baseline, not of machine learning searches in general. \citet{jacobsonbell2025} applied unsupervised clustering to turboSETI hits to remove interference, but their method operates on the output of a narrowband search and so inherits its assumption about signal shape. The morphological assumption from the classical pipelines still exists; it has moved from the search algorithm to the training data.

There have been considerable efforts to remove interference from technosignature searches. \citet{pinchuk2022} trained a CNN to identify RFI using direction of origin, and \citet{zhao2026} applied DBSCAN clustering to residual RFI. Both report the fraction of interference removed, and \citeauthor{zhao2026} also check the retention cost, which is the fraction of genuine signals that survive the filter: of 294 injected simulated signals, none were lost.

That check matters, since a filter that removes 99\% of interference cannot be distinguished, on that figure alone, from one that vetoes everything. However, the simulated signals used for it are single-frequency emissions of the kind the search already assumes. Retention is therefore measured only for the expected morphology, and the cost for any other morphology is unknown.

Fixing a trained model without retraining has its own literature present. \citet{hendrycks2017} began it with a simple observation that an unfamiliar input tends to produce a lower peak confidence. ODIN \citep{liang2018} built on this by adding temperature scaling and input perturbation. \citet{lee2018} took a different path and measured distance from the training distribution in feature space. \citet{liu2020} replaced the softmax score with an energy score. Separately, \citet{guo2017} showed that trained networks are systematically overconfident, and \citet{ovadia2019} tested these corrections under distribution shift.

None of the SETI systems cited here applies any of this. The filter in this paper belongs to the same family of post-hoc correction, but it uses the cadence structure of the observation rather than the model's own output. To my knowledge, neither approach has been tested on cadence data.

This paper addresses the three gaps identified above. A total of 15,000 injections were placed across ten morphologies and five amplitudes into real Green Bank backgrounds, and evaluated against a detector trained once and then frozen, at a threshold fixed before any injected data was scored. Both the fraction of interference removed and the cost in recall are reported for the post-hoc filter, with recall measured on a morphology outside the narrowband assumption. The filter's success criteria were recorded before it was run, so that the outcome could not be reinterpreted afterwards.

\section{Methods}
\subsection{Data}
\begin{figure}[htbp]
\centering
\includegraphics[width=0.98\textwidth]{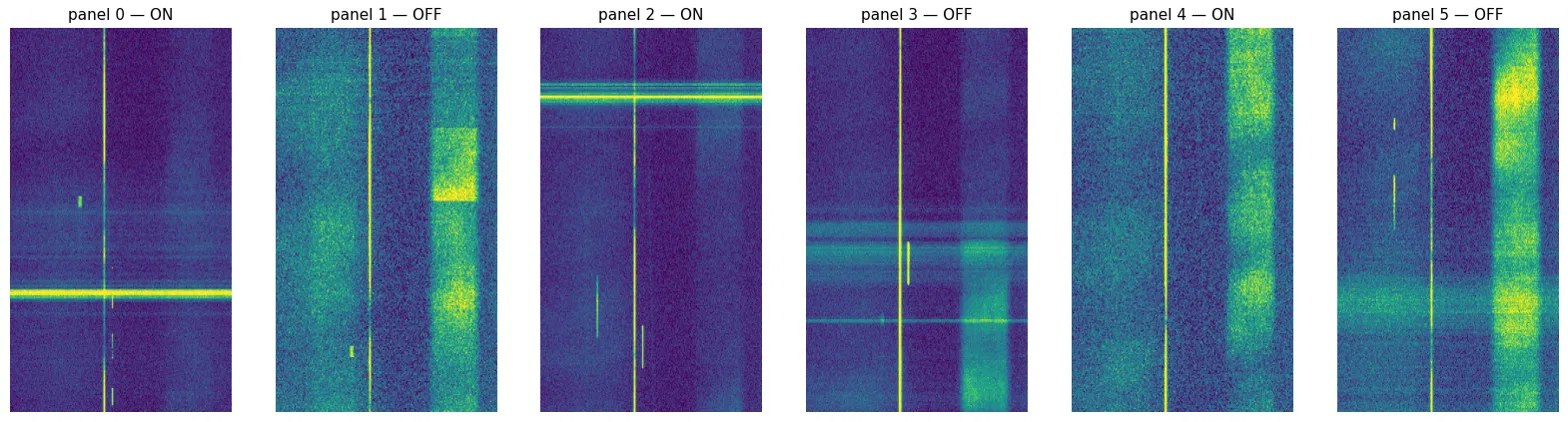}
\caption{An example cadence from the dataset. Panels 0, 2 and 4 are ON (telescope pointed at the target); panels 1, 3 and 5 are OFF. The bright vertical line is present in every panel and is therefore interference.}
\label{fig:cadence}
\end{figure}

Data used is sourced from a Kaggle competition based on SETI Breakthrough Listen \citep{bl2021}; I used the relaunched data release, not the earlier one withdrawn because of a data leak. It includes samples with 6 panels of telescope recordings (273 $\times$ 256 each) and labels marking the presence of a synthetic signal. Panels 0, 2, 4 are pointed towards the target star hence labelled ON, while panels 1, 3, 5 are pointed away and hence labelled OFF (Figure~\ref{fig:cadence}). In each panel, the vertical axis represents time while the horizontal axis represents frequency. A genuine signal can only appear in an ON panel, because it is when the telescope is pointed towards the target star, while interference can appear in all 6 as it comes from Earth, and reaches the telescope no matter where it is pointed.

\subsection{The detector}
The CNN used is EfficientNet-B0 \citep{tan2019}, preceded by a learned resizer \citep{talebi2021}, following a public baseline notebook from the same Kaggle competition \citep{tucker2021}. I trained it once on the competition's training data and then froze all parameters, so every experiment in this paper uses identical weights and no result can reflect a change to the model between experiments. Training used 15 epochs, a fixed random seed of 42 and the notebook's gradual warm-up learning-rate schedule, with fold 0 of five held out for validation; the checkpoint with the best validation AUC was kept, so the AUC below and the threshold derived from the same fold (Section 3.4) are slightly optimistic. The three ON panels are stacked into an 819 $\times$ 256 array, transposed so that frequency runs vertically, and resized to 512 $\times$ 512 before entering the network. The CNN only reads the ON panels, and panels 1, 3, 5 are discarded before the model sees them. On the held-out validation fold it has an Area Under the ROC Curve (AUC) of 0.8545, which signifies how often a model gives a higher score to an image with a signal than to one without it. An AUC score of 0.8545 shows the detector is competent on the kind of data it was trained for, so the failures reported in this paper occur in a working model rather than a broken one.

\subsection{Injection}
I injected synthetic signals into actual telescope backgrounds rather than simulated noise, so that my results cannot be dismissed as artefacts of noise simulators.

I used ten signal families, divided into two groups of five. Direction 1 consists of genuine signal morphologies, shapes a real transmitter might produce. Direction 2 consists of interference morphologies, shapes RFI from Earth produces.

\begin{table}[htbp]
\centering
\small
\caption{The ten injected signal families. Lines are Gaussian in frequency with $\sigma = 1.2$ channels (ch); a panel is 256 channels by 273 time samples.}
\label{tab:families}
\begin{tabular}{ll>{\raggedright\arraybackslash}p{4.3cm}>{\raggedright\arraybackslash}p{5.2cm}}
\toprule
ID & Name & Description & Parameters \\
\midrule
\multicolumn{4}{l}{\textbf{Direction 1} --- \textit{genuine signal morphologies}} \\
D1-1 & extreme drift & a straight line with a steep tilt & linear drift of 60 ch across the panel \\
D1-2 & quadratic & a line that curves smoothly in one direction & quadratic path, 60 ch of curvature across the panel \\
D1-3 & S-shaped & a line that oscillates from side to side & sinusoid of $\pm$25 ch, one cycle per panel \\
D1-4 & pulsed & a line broken into separate segments & drift of 5 ch; on for 35\% of each 55-sample period \\
D1-5 & dispersed burst & a line that sweeps and curves sharply & broadband; arrival at sample $30 + 180(f/256)^2$, $\sigma = 6$ samples \\
\addlinespace
\multicolumn{4}{l}{\textbf{Direction 2} --- \textit{interference morphologies}} \\
D2-1 & zero drift & a perfectly vertical line with no tilt & no drift \\
D2-2 & comb & several parallel lines at once & four tones 18 ch apart, no drift \\
D2-3 & intermittent & a line that appears and disappears irregularly & no drift; on for 40\% of each 40-sample period \\
D2-4 & broadband burst & a wide horizontal smear & all channels; Gaussian in time, $\sigma = 3$ samples \\
D2-5 & high drift & a thin line with a steep tilt & linear drift of 45 ch across the panel \\
\bottomrule
\end{tabular}
\end{table}

\begin{figure}[htbp]
\centering
\includegraphics[width=0.98\textwidth]{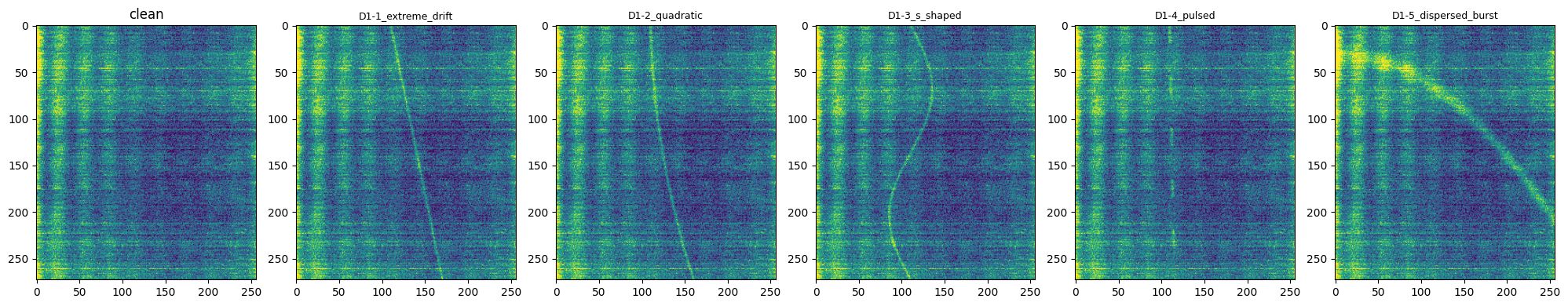}
\caption{Examples of the five Direction 1 morphologies, each beside the clean background.}
\label{fig:d1}
\end{figure}

\begin{figure}[htbp]
\centering
\includegraphics[width=0.98\textwidth]{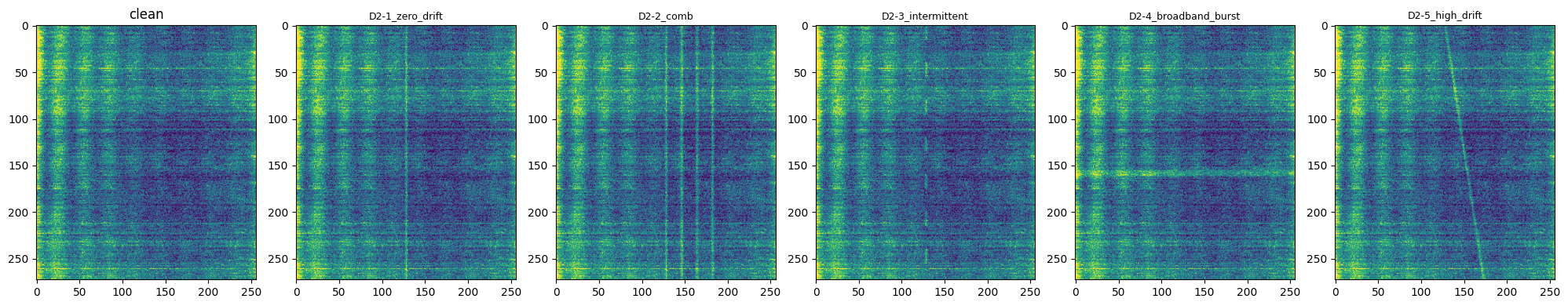}
\caption{Examples of the five Direction 2 morphologies, each beside the clean background.}
\label{fig:d2}
\end{figure}

Examples of each morphology are shown in Figures~\ref{fig:d1} and~\ref{fig:d2}.

Direction 1 signals were injected into the ON panels only, as a genuine signal comes from a star and can only be present while the telescope is pointed towards it. Direction 2 interference was injected into all six panels as RFI comes from Earth and is present regardless of pointing. Because the detector reads only panels 0, 2 and 4, what is placed in panels 1, 3 and 5 cannot change its score: to the detector, a Direction 2 injection differs from a Direction 1 injection only in shape. D1-1 and D2-5, for example, are both straight lines, drifting 60 and 45 frequency channels across a panel respectively. The starting channel of each injection was drawn at random between channels 64 and 192 and recorded in the manifest, and the parameters of each family are given in Table 1.

Signals with five different amplitudes were injected into each image: 0.25, 0.5, 1.0, 2.0, 4.0. The competition data are distributed already normalised to approximately zero mean and unit standard deviation, so amplitude, the peak of the injected line in each time row, is expressed in units of the noise standard deviation. I calibrated this scale with a narrowband needle of the kind in the training data, drifting 3 channels across a panel: at an amplitude of 0.5 its recall was 0.590 (95\% CI 0.492--0.681, n = 100), consistent with the baseline recall, so 0.5 is the amplitude at which an injection of the familiar kind reproduces the detector's baseline performance. An amplitude of 1.0 meant the signal was as bright as the typical noise in the image, while an amplitude of 4.0 meant it was 4 times brighter than the typical noise in the image.

A total of 300 samples were generated for each family per amplitude. Five amplitudes $\times$ ten families led to a total of 15,000 injections.

Details of every injection were recorded in a manifest file: amplitude, source image, family, injection position and the random seed. This means every injection can be regenerated exactly, making the entire experiment reproducible. The Kaggle data carry no frequency or time metadata, so drifts are given in channels per panel rather than in Hz s$^{-1}$.

Three simplifications should be noted. Each ON panel received the same injected path, starting at the same channel, rather than a signal whose drift continues across the cadence. All 15,000 injections used the same 300 backgrounds, drawn from signal-free samples in the competition's training set, about four in five of which come from the folds the detector was trained on; the same 300 backgrounds also served as the clean controls (Section 3.4) and calibrated the cadence filter (Section 3.6). And the competition normalised each sample after injecting its signals, whereas my injections were added to already-normalised backgrounds and not renormalised. Because every family used the same backgrounds, each injected score can be paired with the score of its own clean background.

\subsection{Scoring}
The model gives every image a number between 0 and 1, signifying how confident the model is that a signal exists. To interpret a model's confidence into an either yes or no, a threshold was required; above it means detected and below it means missed. To find a threshold, I scored the held-out validation fold of the competition data and set the threshold at the level below which 99\% of its 10,800 signal-free samples sat, which came out to be 0.1309. At that threshold, 663 of the 1,200 validation samples containing a signal were detected, a baseline recall of 0.5525 (95\% CI 0.524--0.580) on the kind of signal the detector was trained on. As a check, I scored 300 clean control images, real telescope noise with nothing injected, and 4 of them crossed the threshold: a false positive rate of 0.0133 (95\% CI 0.005--0.034), consistent with the 1\% I designed it for.

I fixed the threshold at 0.1309 before scoring any injected data, and every family and amplitude was evaluated against the same value. This resulted in fair tests, which allowed differences between families to reflect the model's behaviour, not the way I scored them.

For Direction 1, I measured recall: the fraction of injected signals that scored above the threshold. For Direction 2, I measured false positive rate: the fraction of injected interference that scored above the threshold. Each rate is computed on 300 injections, and error bars in the figures are 95\% Wilson confidence intervals.

The reason for not using accuracy as a metric is that it can turn out quite misleading. In SETI data, almost every observation contains no signal. A model that rejected everything would score near-perfect accuracy and detect nothing. Recall and false positive rate avoid this as they look at both: images with signals, and empty images separately rather than mixing the two groups together.

\subsection{Attention analysis}
To find the reason behind failure for certain families, I used Grad-CAM \citep{selvaraju2017}. Grad-CAM produces a heatmap over the image showing the regions which influenced the model's score the most. I implemented it myself rather than using a library. Grad-CAM was computed at an intermediate layer, where each cell of the 16 $\times$ 16 map covers 16 frequency channels; the final convolutional layer's 8 $\times$ 8 map covers 32 channels per cell, too coarse to localise a line one or two channels wide.

To isolate the effects of injection in the image, I first computed the heatmap for a clean image, then the same image with an injection, and subtracted. The background therefore cancels, and what remains is exactly what the injection changed.

I used two measures on this difference map. The first is its total magnitude, the sum of its absolute values, which gives the size of the response to the injection. The second is the on-path test: the mean of the map along the injected signal's path, taken from its generator rather than read off the figure, divided by the mean of the same map along other samples' paths. A ratio above 1.3, set before the test was run, counts as attention that follows the signal. Paired comparisons in this analysis were tested with the Wilcoxon signed-rank test.

The analysis was run on a cohort of 72 samples, made up of six groups of twelve: Q1\_fool, Q1c\_true\_det, Q2\_bright\_hit, Q2c\_bright\_mis, Q3\_pulsed and Q3\_dispersed. I locked them before the analysis, using a stated rule, so I could not pick what looked the best at a later phase. Each group was drawn from one family at one amplitude. Q1\_fool and Q1c\_true\_det are the twelve highest-scoring detections of D2-5 and D1-2 at an amplitude of 0.5. Q2\_bright\_hit is the twelve highest-scoring detections of D1-3, and Q2c\_bright\_mis the twelve lowest-scoring misses of D1-1, both at an amplitude of 4.0. Q3\_pulsed and Q3\_dispersed are twelve injections of D1-4 and D1-5 drawn at random with a fixed seed, at an amplitude of 4.0.

The heatmap shows where the model looked but not if it is causal. Hence, I erased the region Grad-CAM marked as important, and re-scored to check whether the score drops. As a control, I erased random regions of the same size. I set the criterion in advance: erasing the three highest-ranked regions should remove at least half of the gap between the injected and clean scores, in at least 9 of 12 samples. If it does, it means the attention is indeed causal and not just decorative.

Grad-CAM can sometimes produce convincing-looking maps even from a randomised model. To counter this, I ran the randomisation check of \citet{adebayo2018}, comparing the attention maps produced by the trained model against those produced after its weights were randomised.

\subsection{The cadence filter}
The model does not read the OFF panels, so I tested whether the missing check could be added afterwards. The filter runs after the model: the model flags an image, and the filter then inspects the 6 panels and can veto the decision. No retraining was involved.

\begin{figure}[htbp]
\centering
\includegraphics[width=0.5\textwidth]{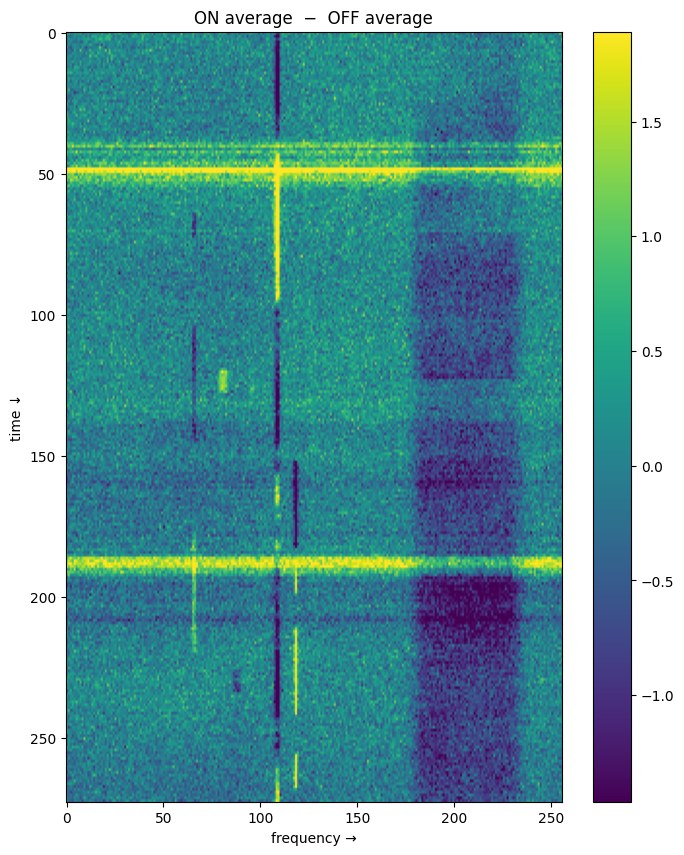}
\caption{Average of the ON panels minus average of the OFF panels for the cadence in Figure~\ref{fig:cadence}.}
\label{fig:onoff}
\end{figure}

The filter measures ON-OFF contrast, the difference in energy between the ON and OFF panels (Figure~\ref{fig:onoff}). Precisely, it averages each frequency channel over time and over the three ON panels, does the same for the three OFF panels, subtracts the two profiles, smooths the result over 9 channels, and takes its maximum, S. For a genuine signal, the difference between ON and OFF should be noticeable because the signal is only present in an ON panel; hence it is kept. In case of RFI however, the difference should appear smaller, as interference is present in both panels and partly cancels; hence it is vetoed. Because the statistic averages over time before comparing ON and OFF, a drifting line is spread across every channel it crosses: a line drifting D channels contributes roughly a fraction 1.2$\sqrt{2\pi}$/D of its amplitude to each channel's average, about 0.05 for the 60-channel drift of D1-1. The statistic is therefore far less sensitive to drifting signals than a comparison made along the signal's path, as de-Doppler pipelines do.

I calibrated the threshold on 300 clean backgrounds and set it to their 99th percentile, 1.0266. Detections scoring below this were vetoed. The filter was applied to all 2,277 detections the model flagged.

Before running the filter, I recorded what would count as success: false positive rate had to drop, and the cost in recall had to stay within 10 percentage points. Both conditions had to hold, not either.

I discarded an earlier version of the filter before any results were measured. Its sanity check failed because the peak-finder was locking onto background interference rather than the injected signal, so it was replaced by the design above.

\section{Results}
\subsection{Direction 1 --- missed detections}
Before running Direction 1, I recorded the prediction that recall would decline as morphology grew more distant from the training distribution, in the order D1-1, D1-2, D1-3, D1-4, D1-5. The ordering was an assumption rather than a measurement: I did not characterise the morphologies in the training data, which the competition states are not all straight lines and are not always present for the whole of each ON observation \citep{bl2021}. The results did not follow this prediction.

\begin{figure}[htbp]
\centering
\includegraphics[width=0.98\textwidth]{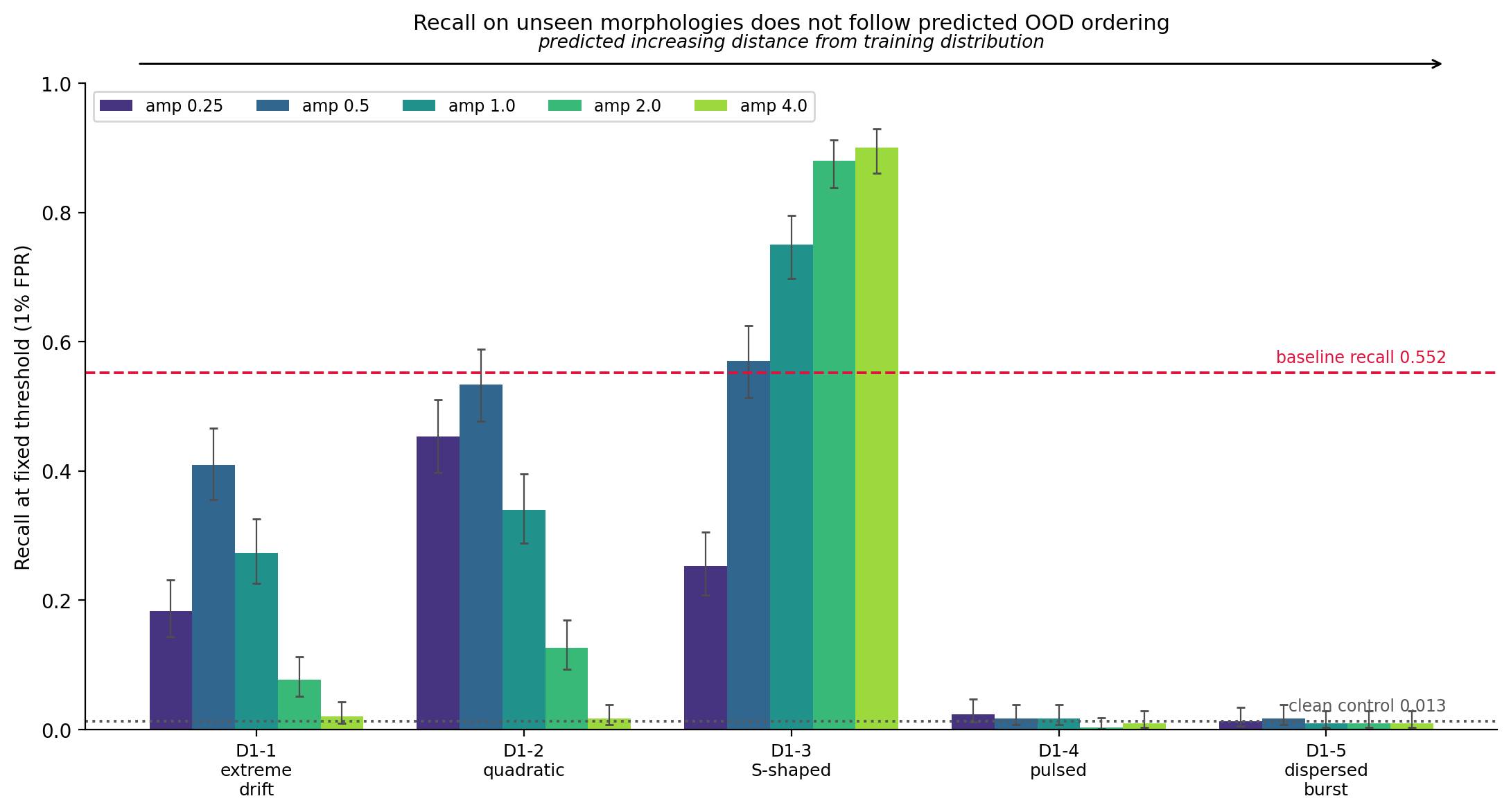}
\caption{Recall at the fixed threshold for each Direction 1 family and amplitude. The dashed line marks baseline recall (0.5525); the dotted line marks the clean control rate (0.0133).}
\label{fig:recallfam}
\end{figure}

Recall varied sharply across the five families (Figure~\ref{fig:recallfam}). D1-3 S-shaped achieved the highest recall at 0.900, despite being predicted to sit in the middle of that ordering. Only the far end of the ordering held: D1-4 pulsed and D1-5 dispersed burst stayed near the clean control rate of 0.0133 at every amplitude. Baseline recall on the signal type the model was built for is 0.5525.

\begin{figure}[htbp]
\centering
\includegraphics[width=0.98\textwidth]{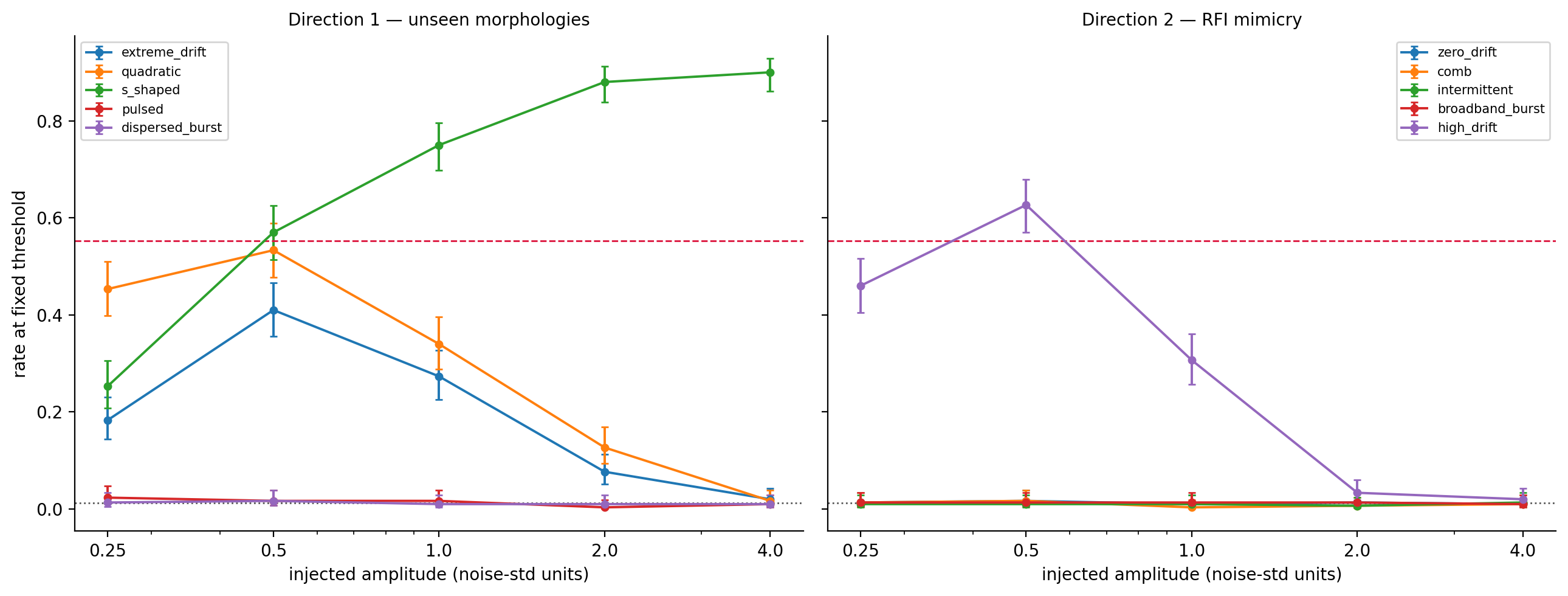}
\caption{Detection rate against injected amplitude. Left: recall for Direction 1. Right: false positive rate for Direction 2.}
\label{fig:rateamp}
\end{figure}

Recall did not increase with amplitude for every family (Figure~\ref{fig:rateamp}). D1-1 and D1-2 peaked at an amplitude of 0.5, reaching 0.410 and 0.533 respectively, then fell to 0.020 and 0.017 respectively at an amplitude of 4.0. D1-3 rose across the range from 0.253 at an amplitude of 0.25, and gradually increased to 0.900 at an amplitude of 4.0. The fall is not specific to unfamiliar morphologies: the calibration needle, of the kind the detector was trained on, dropped from a recall of 0.590 at an amplitude of 0.5 to 0.090 at 4.0 (n = 100 each).

Pairing each injection with its own clean background shows the same fall in the scores themselves. For D1-1, the median increase in score over the clean background was 0.049 at an amplitude of 0.5 and 0.022 at 4.0; at 4.0 the injected score still exceeded the clean one in 94\% of samples. For D1-4 and D1-5, the median change was within 0.003 of zero at every amplitude.

As a sanity check that these differences reflect shape rather than injected energy, I computed the energy each family deposits. At a matched amplitude, D1-1, D1-2 and D1-3 deposit the same energy, as expected for lines of equal width and amplitude. D1-4 deposits 0.366$\times$ that of D1-1; D1-5 deposits 4.68$\times$ that of D1-1.

\subsection{Direction 2 --- false alarms}
\begin{figure}[htbp]
\centering
\includegraphics[width=0.98\textwidth]{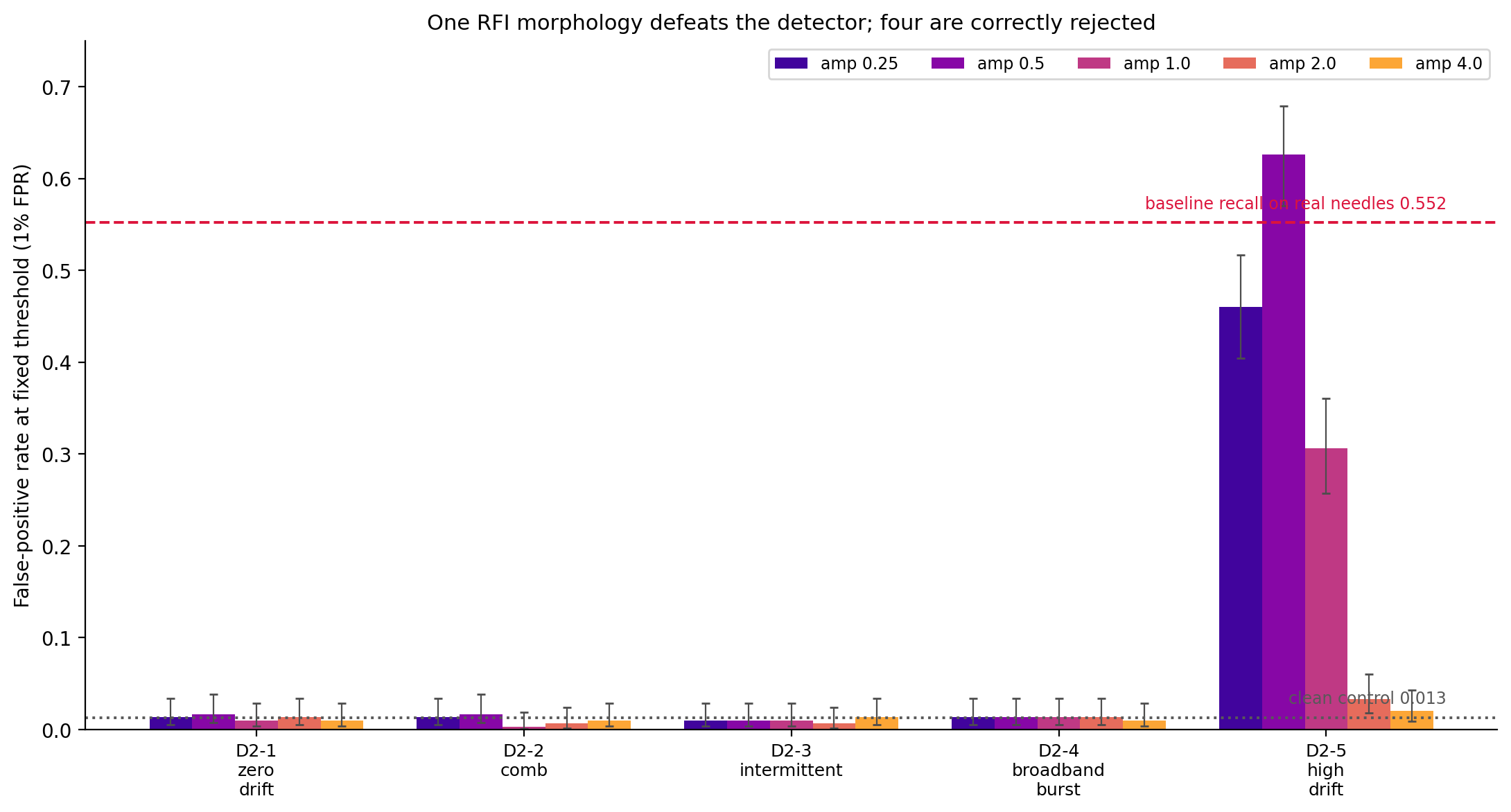}
\caption{False positive rate at the fixed threshold for each Direction 2 family and amplitude. The figure's label `real needles' refers to the competition's labelled needles, which are synthetic.}
\label{fig:fprfam}
\end{figure}

Four of the five interference families were rejected, the correct outcome for interference (Figure~\ref{fig:fprfam}). D2-1 zero drift, D2-2 comb, D2-3 intermittent and D2-4 broadband burst all remained near the clean control rate of 0.0133 at every amplitude.

D2-5 however was not rejected. Its false positive rate was 0.460 at an amplitude of 0.25, rose to 0.627 at an amplitude of 0.5 and then dropped to 0.307, 0.033 and 0.020 at amplitudes of 1.0, 2.0 and 4.0 respectively.

At its peak, D2-5's false positive rate of 0.627 was nominally above the model's baseline recall of 0.5525 on genuine signals (two-proportion test, p = 0.018). Because 0.627 is the highest of five amplitudes, the difference is not significant once that selection is accounted for (p $\approx$ 0.09), and the two rates are best described as comparable. A like-for-like comparison makes the same point: D1-1, a genuine signal that is also a straight line but drifts 60 channels rather than 45, reached a recall of 0.410 at the same amplitude.

The false positive rate for D2-5 followed the same non-monotonic pattern seen in D1-1 and D1-2 in Direction 1, peaking at an amplitude of 0.5 and falling to its lowest at an amplitude of 4.0 (Figure~\ref{fig:rateamp}).

\subsection{Attention analysis results}
\begin{figure}[htbp]
\centering
\begin{minipage}{0.49\textwidth}\centering\includegraphics[width=\linewidth]{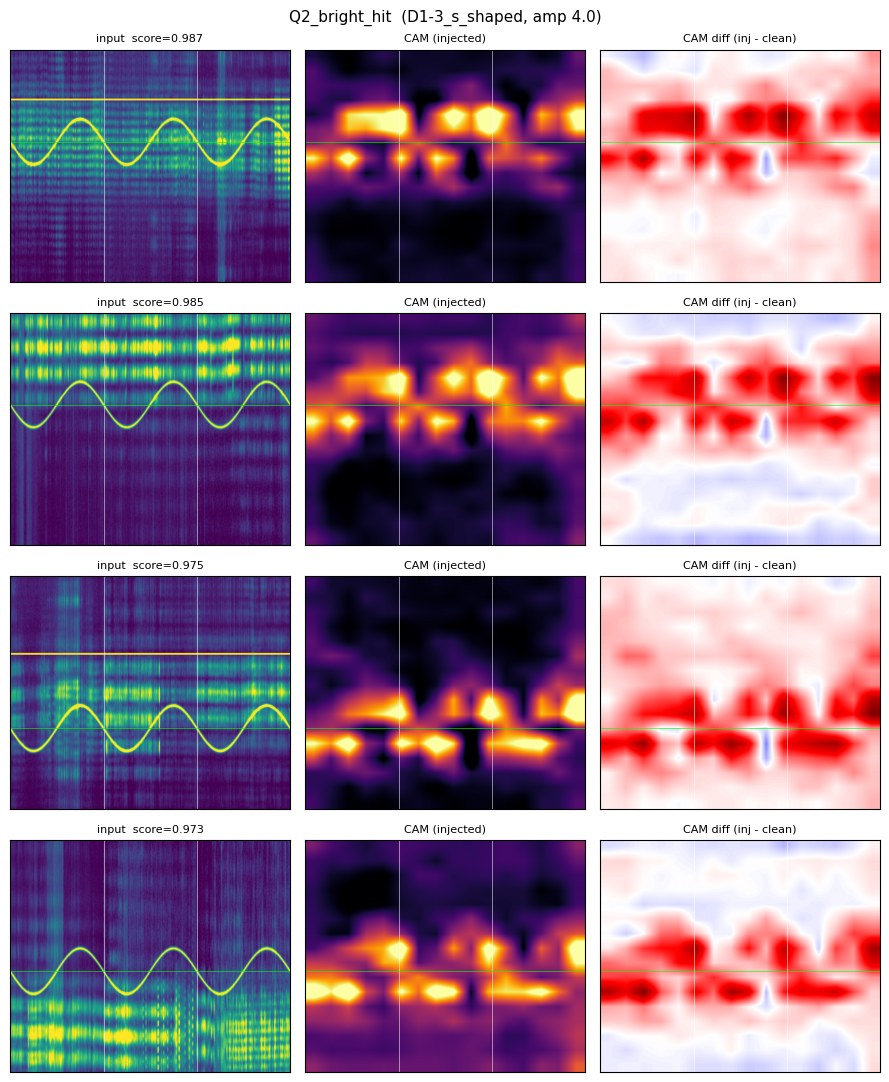}\\(a)\end{minipage}\hfill
\begin{minipage}{0.49\textwidth}\centering\includegraphics[width=\linewidth]{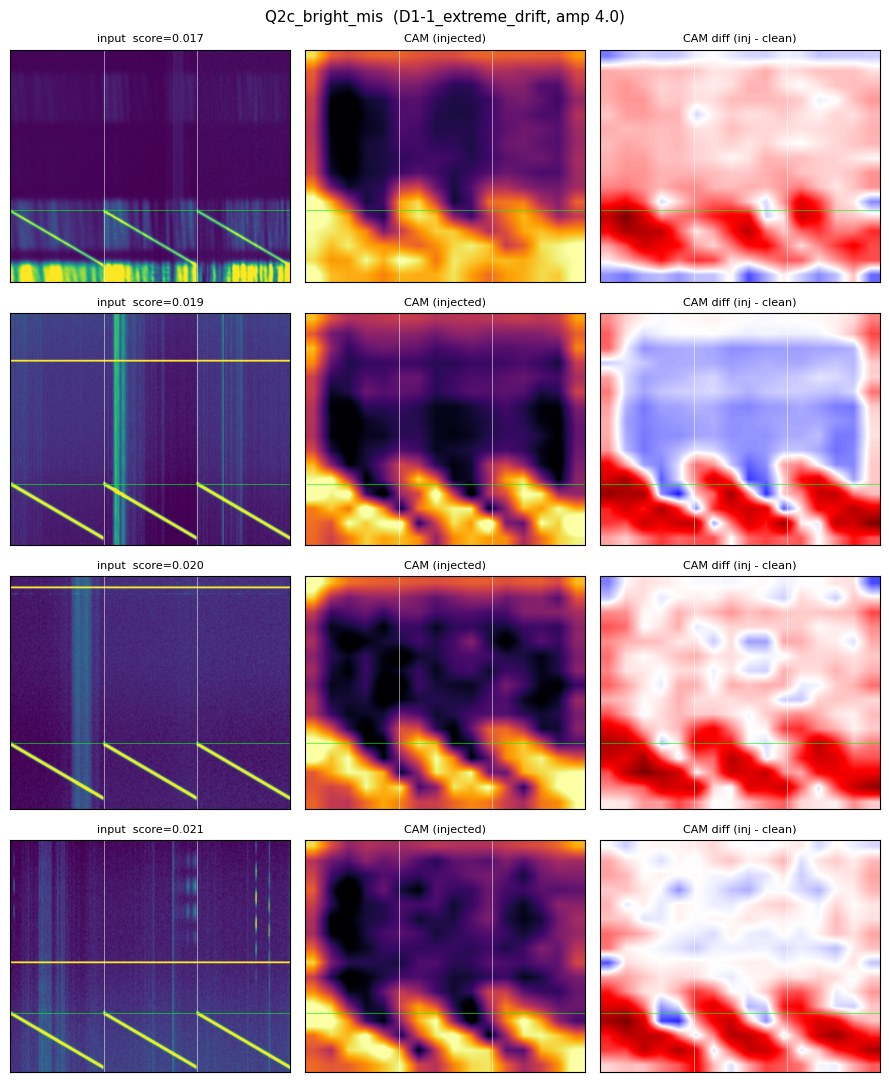}\\(b)\end{minipage}
\caption{Grad-CAM autopsy at an amplitude of 4.0. (a) Q2\_bright\_hit: D1-3 S-shaped, detected. (b) Q2c\_bright\_mis: D1-1 extreme drift, missed. Columns: input, Grad-CAM map, and CAM difference (injected minus clean). Inputs are shown as the detector sees them: the three ON panels side by side, with frequency running vertically; white lines mark panel boundaries and the green line the injection's starting channel.}
\label{fig:q2}
\end{figure}

To examine what the model responded to, I compared two groups of twelve injections at the same amplitude of 4.0: D1-3 S-shaped detections, scoring up to 0.987, and D1-1 extreme drift misses, scoring as low as 0.017 (Figure~\ref{fig:q2}).

For the missed D1-1 injections, the CAM difference shows attention along the injected line. This reading is visual: the on-path test was inconclusive for D1-1 (ratio 1.06, p = 0.18), because all twelve paths start between channels 165 and 191, within about 1.6 cells of the attention map, so other samples' paths are not a valid null.

D1-3 produced 2.9$\times$ the total CAM perturbation that D1-1 did (1.534 against 0.534). I tested two alternative explanations for this difference, path curvature and response concentration, and rejected both: curvature showed no effect, and concentration differed only marginally, with the top 10\% of the map holding 0.414 of the response for D1-3 against 0.358 for D1-1. Because the two groups were selected as the highest-scoring detections and the lowest-scoring misses, the difference in total perturbation partly reflects that selection.

\begin{figure}[htbp]
\centering
\begin{minipage}{0.49\textwidth}\centering\includegraphics[width=\linewidth]{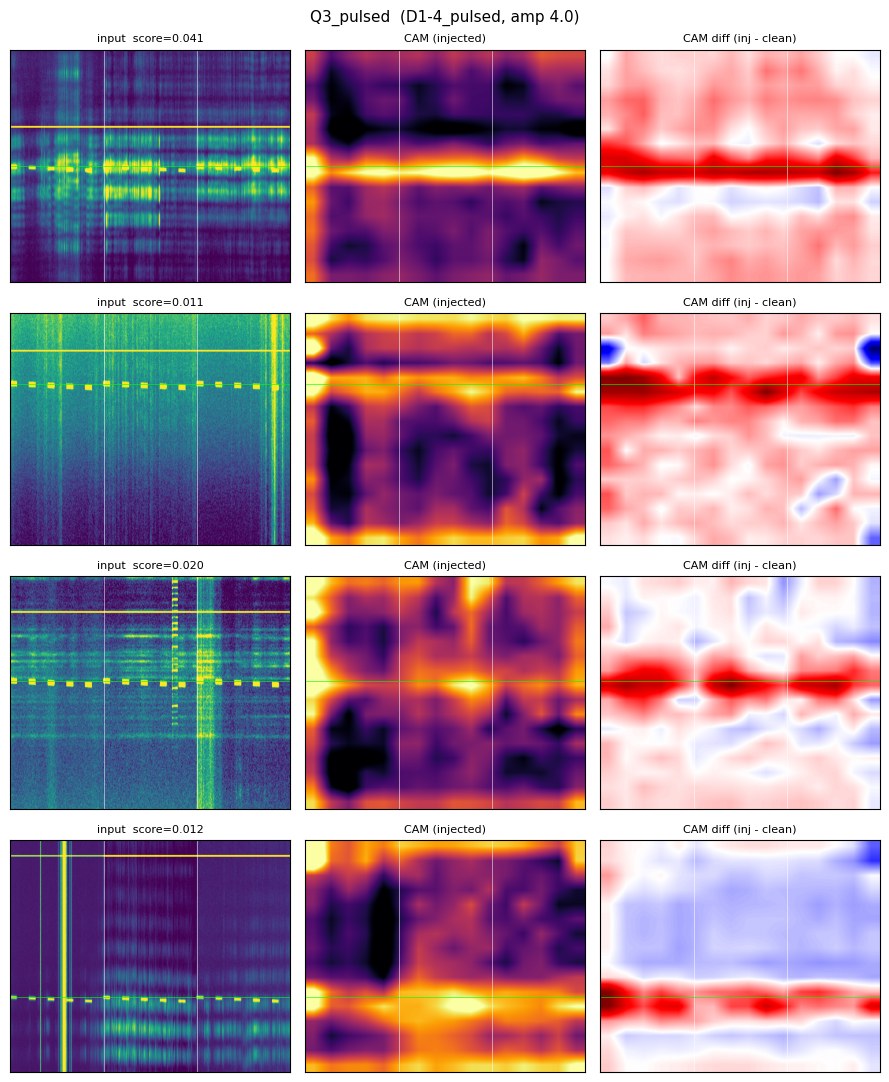}\\(a)\end{minipage}\hfill
\begin{minipage}{0.49\textwidth}\centering\includegraphics[width=\linewidth]{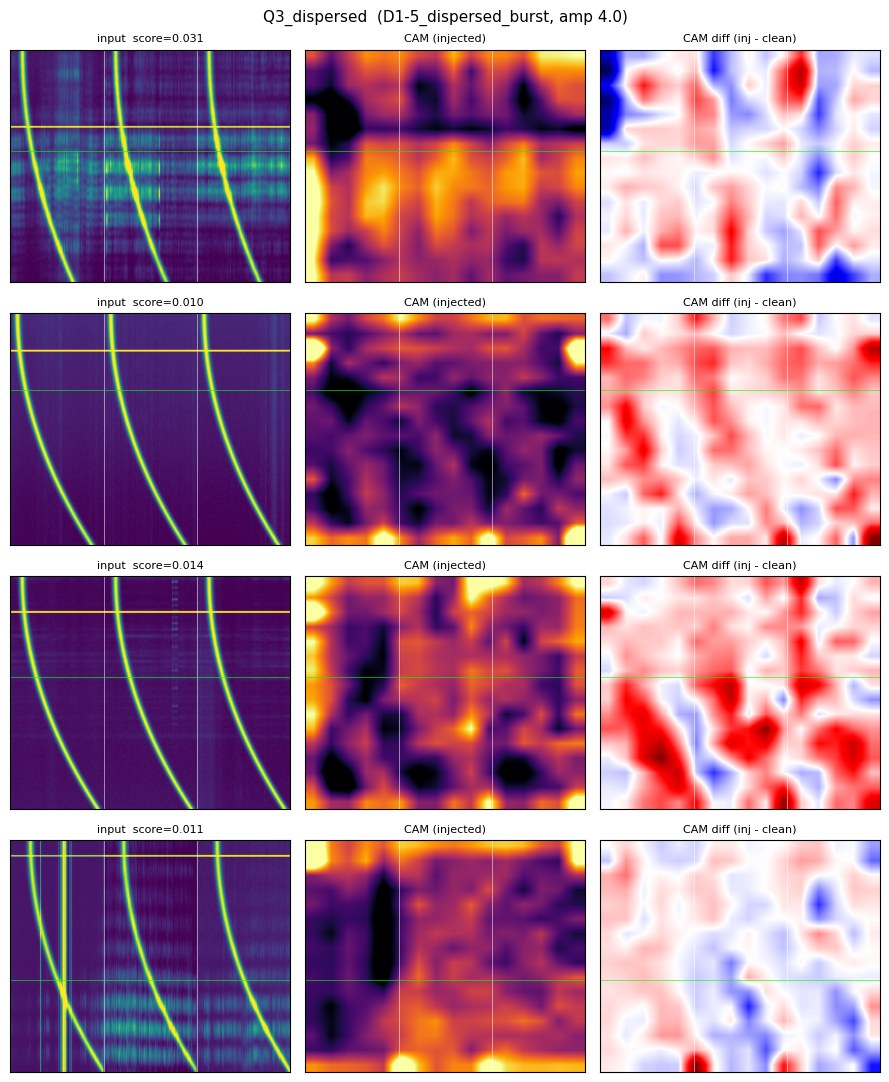}\\(b)\end{minipage}
\caption{Grad-CAM autopsy at an amplitude of 4.0. (a) Q3\_pulsed: D1-4. (b) Q3\_dispersed: D1-5. Both missed. Layout as in Figure~\ref{fig:q2}.}
\label{fig:q3}
\end{figure}

D1-4 pulsed showed structured attention along the injected path, with an on-path ratio of 2.81$\times$ (above 1 in 12 of 12 samples, p = 0.0005). D1-5 dispersed burst did not meet the 1.3 threshold, at 1.15$\times$ (11 of 12, p = 0.003), a consistent but small effect, despite depositing about 13$\times$ the energy of D1-4 (Figure~\ref{fig:q3}).

\begin{figure}[htbp]
\centering
\begin{minipage}{0.49\textwidth}\centering\includegraphics[width=\linewidth]{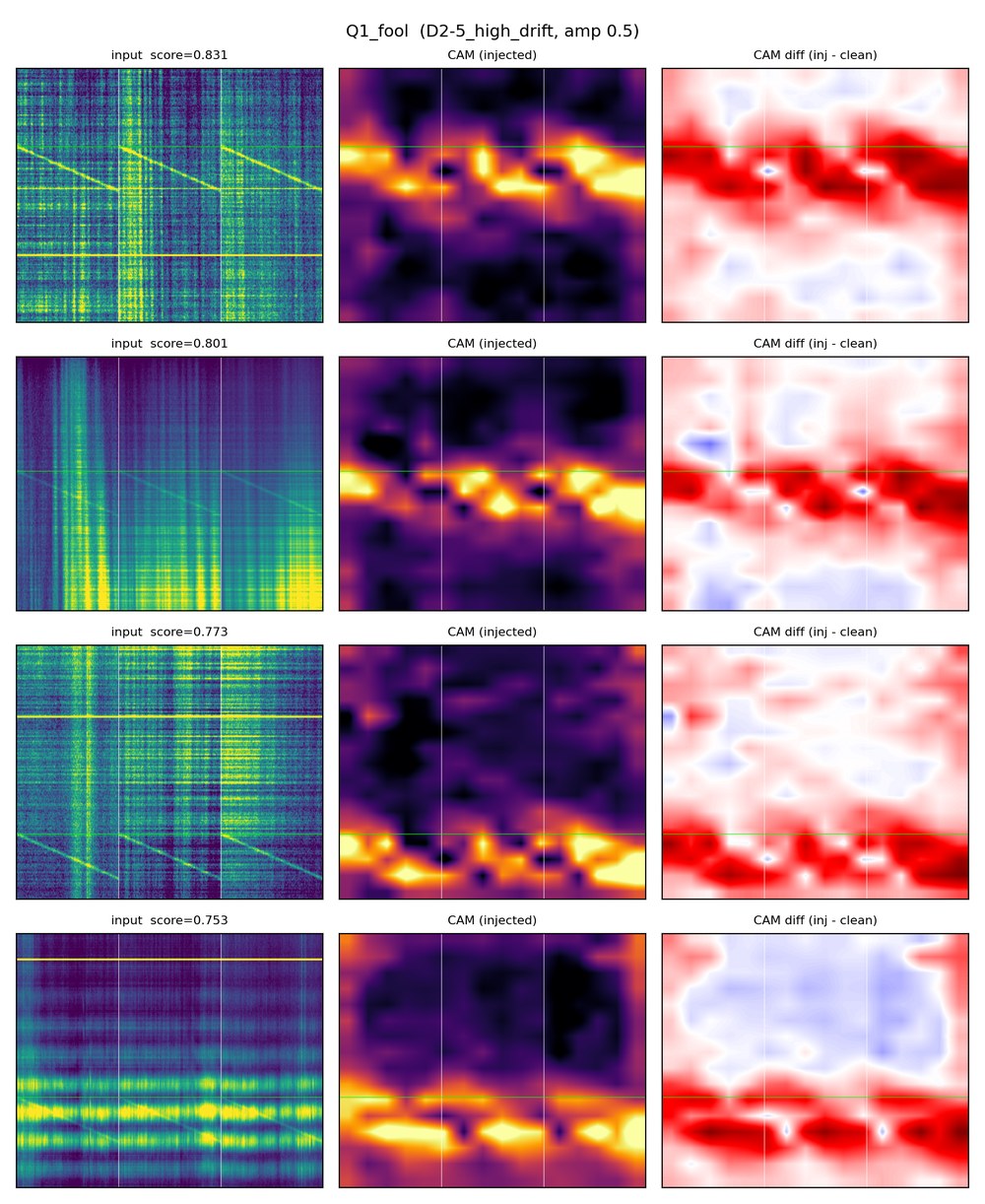}\\(a)\end{minipage}\hfill
\begin{minipage}{0.49\textwidth}\centering\includegraphics[width=\linewidth]{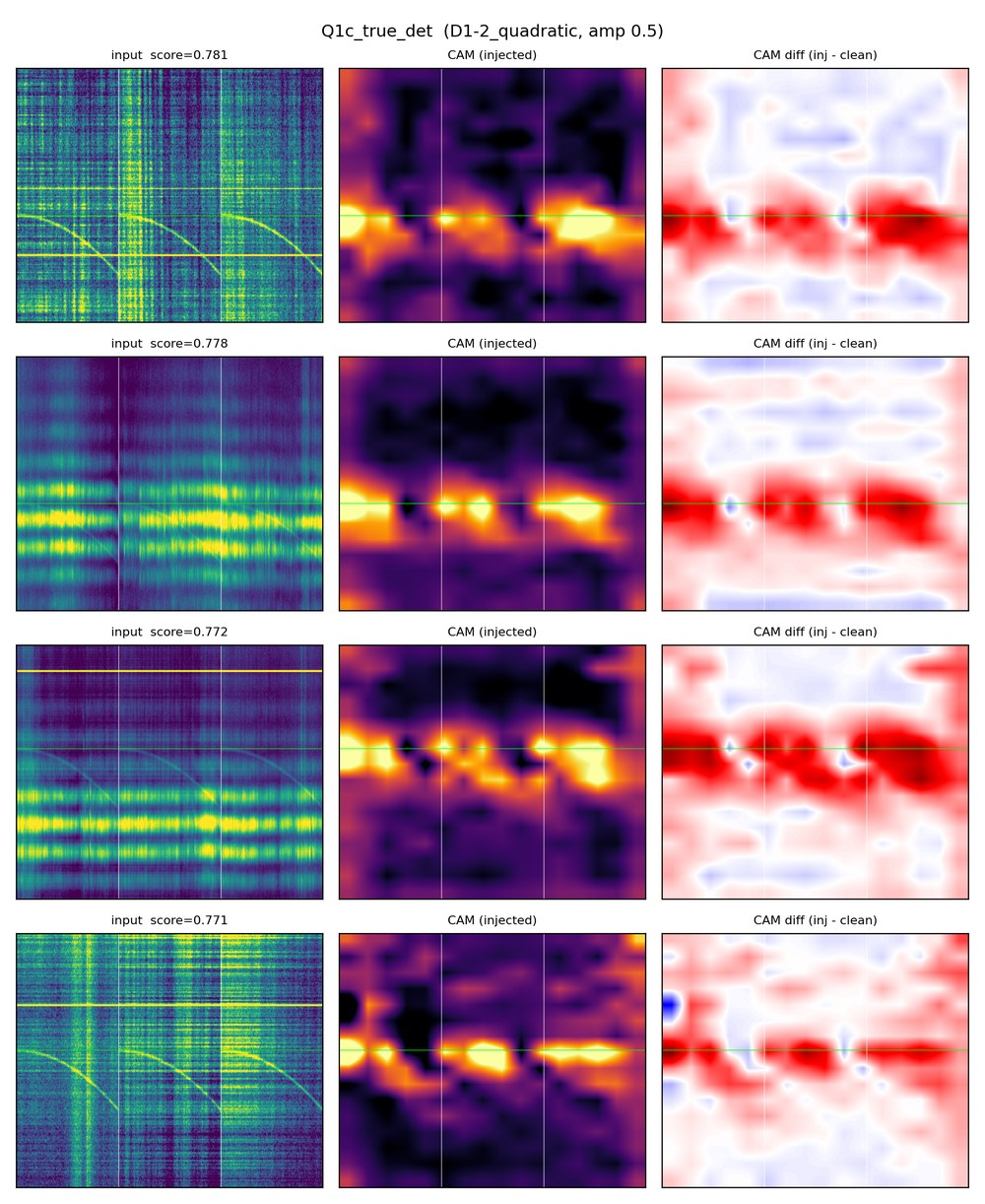}\\(b)\end{minipage}
\caption{Grad-CAM autopsy at an amplitude of 0.5 and matched scores. (a) Q1\_fool: D2-5 high drift, a false alarm. (b) Q1c\_true\_det: D1-2 quadratic, a true detection. Layout as in Figure~\ref{fig:q2}.}
\label{fig:q1}
\end{figure}

The Q1 pair compares a false alarm with a true detection at matched confidence. Both were taken at an amplitude of 0.5: the D2-5 false alarms had a mean score of 0.710, and the D1-2 true detections 0.695 (Figure~\ref{fig:q1}). In both groups, attention sat along the injected line. The deletion test gave the same pattern for each: in a post-hoc look at six regions, erasing the six regions Grad-CAM ranked highest removed 0.85 and 0.68 of the gap between injected and clean scores respectively, while erasing random regions of the same size removed almost none.

The deletion test did not meet its pre-specified criterion for any detected group: erasing three regions removed half the gap in only 3, 4 and 7 of 12 samples for Q1\_fool, Q1c\_true\_det and Q2\_bright\_hit. In a post-hoc look at six regions, Grad-CAM-ranked erasure had removed 0.85, 0.68 and 0.75 of the gap, against 0.00, 0.01 and 0.12 for random erasure. For the missed groups, erasing the highest-ranked regions raised the score more than erasing random regions, in 10 of 12 samples for D1-1 (p = 0.007) and for D1-4 (p = 0.002), but not for D1-5 (p = 0.23). The randomisation check gave a rank correlation of +0.029, confirming that the maps come from the trained model rather than the architecture.

The in-band CAM ratio was intended as evidence that the model ignores certain signals, but was retired during this analysis, as it measured dilution rather than absence.

\subsection{The cadence filter}
Applying the cadence filter at an amplitude of 0.5 led to D2-5's false positives falling from 0.627 to 0.007 (2 of 300), a 99\% reduction, which is what the filter was built to do.

However it also removed genuine detections, at the same rate. Across all families and amplitudes it kept 25 of 1,775 genuine detections (1.4\%) and 6 of 502 detections of interference (1.2\%). No Direction 1 family kept a recall above 0.013 at any amplitude; D1-3, for example, fell from 0.900 to 0.013 at an amplitude of 4.0.

Against the pre-specified bar, the false positive rate had to drop and recall had to stay within 10 percentage points. It did pass the first condition, but failed the second by a wide margin. Both had to hold, so the filter fails.

\begin{figure}[htbp]
\centering
\includegraphics[width=0.85\textwidth]{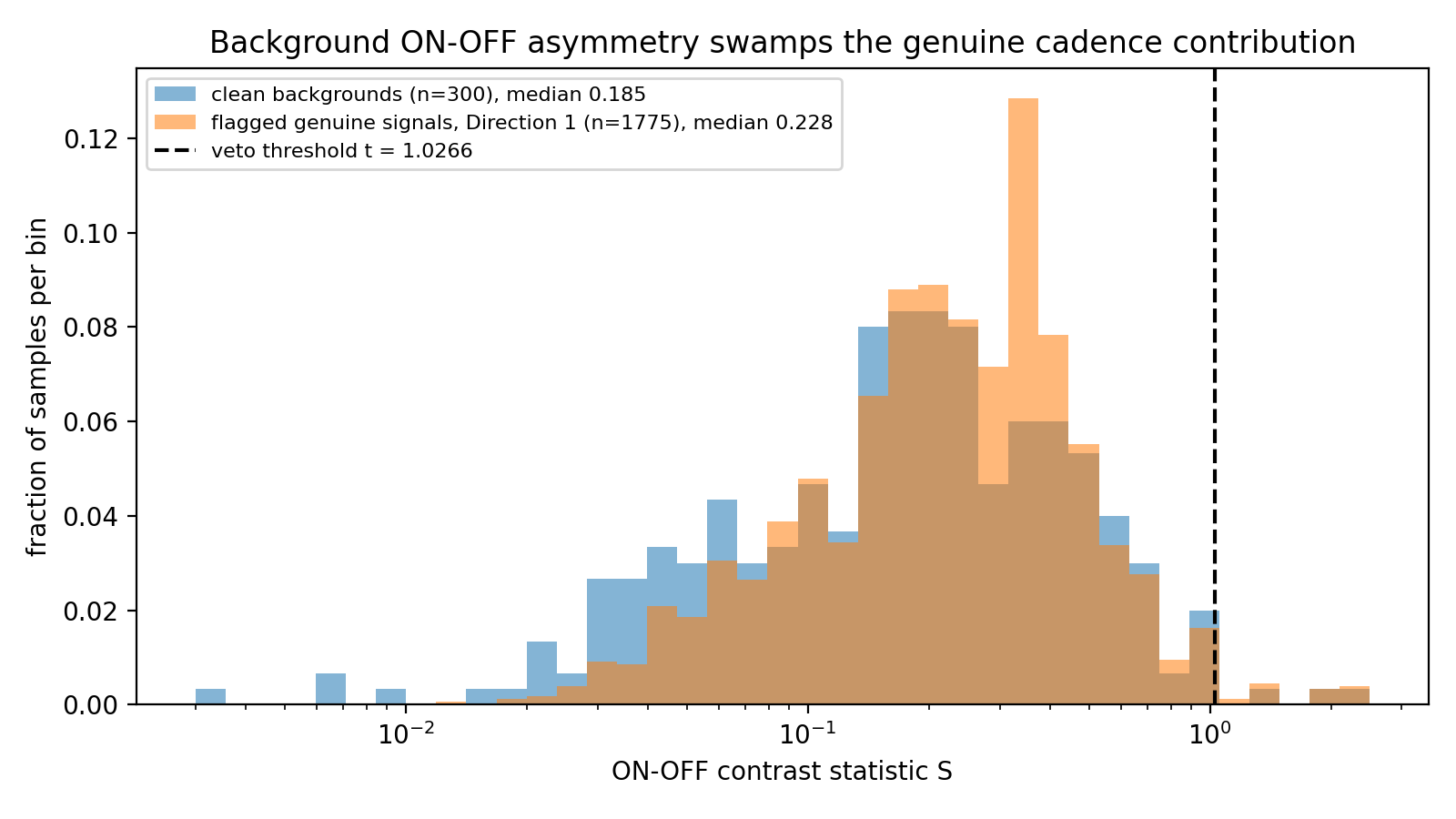}
\caption{Distribution of the ON-OFF contrast statistic S for 300 clean backgrounds and 1,775 flagged genuine signals. The dashed line is the veto threshold, t = 1.0266. Genuine signals are pooled across families; D1-3 contributes 57\% of them.}
\label{fig:phase55}
\end{figure}

To identify why, I measured the ON-OFF asymmetry present in real backgrounds. The median ON-OFF asymmetry was 0.185 in 300 clean backgrounds and 0.228 in 1,775 flagged genuine signals, so a genuine signal shifts the statistic by only about 0.04 (Figure~\ref{fig:phase55}), largely because the time averaging described in Section 3.6 dilutes a drifting line before the comparison.

The statistic also barely separates genuine signals from interference: among flagged detections, it ranks a genuine signal above a D2-5 false alarm only 60\% of the time (AUC 0.60, where 0.5 is chance). Any threshold that vetoes the interference therefore also vetoes the genuine signals.

\section{Discussion}
\subsection{What the ON-only input explains}
The detector reads only ON panels. The single property that separates a genuine candidate from terrestrial interference is presence in OFF panels, and that information is discarded before the model sees the input. The model therefore cannot be asking whether a signal originates from the target star; it can only ask whether the signal matches the learned shape. Matching on appearance rather than on the property that defines the label is an instance of what \citet{geirhos2020} call shortcut learning.

Matching on appearance also explains the Direction 1 results, but this part does not depend on the OFF panels: a detector that read all six panels would still miss a morphology it never learned to recognise. D1-4 and D1-5 remained at the clean control rate at every amplitude. \citet{price2020} confined their analysis to narrow drifting signals and named pulsed signals as future work; the detector tested here does not recover them either.

The highest recall was recorded for D1-3 S-shaped, a morphology predicted to sit mid-way from the training distribution, above the two predicted to be closest. Because the ordering was an assumption and the competition's training signals were not all straight lines, D1-3 may simply have resembled signals the detector was trained on. Its larger attention response (Section 4.3) cannot settle this, because the two groups compared were selected by score. Injected energy does not set the response either: D1-5 deposits about 13$\times$ the energy of D1-4 yet provokes a weaker response, 0.29$\times$ the total CAM perturbation of D1-1 against 0.47$\times$ for D1-4, and this analysis does not explain why.

The ON-only input does explain Direction 2. Interference whose morphology resembles the training distribution is marked as a genuine signal, and is therefore kept. D2-5 reached a false positive rate of 0.627. The attention analysis shows the same thing directly: at matched scores, the model responds to a D2-5 false alarm and a D1-2 true detection in the same way, so its attention does not distinguish between them.

The detector therefore flags this interference about as readily as it detects a genuine signal.

This is not a claim that the detector cannot reject interference at all. D2-1 to D2-4 were all rejected. The claim is narrower: it can reject interference that looks different, but cannot reject interference that looks like a genuine signal.

\subsection{Rejection by appearance, not origin}
Four out of the five interference families were rejected. This might suggest that the detector filters interference, but as established above, the model does not read OFF panels, hence it has no access to the evidence that defines interference.

These families were most likely rejected because their morphologies do not match any one of the shapes of a genuine signal; unlike D2-1 below, this was not tested directly. A comb consists of multiple parallel lines, a broadband burst is a wide horizontal smear and an intermittent signal is a broken flickering line. None of these match a genuine signal's morphology.

This is illustrated most clearly by D2-1 zero drift. A genuine signal from a target star is expected to drift, because Earth is rotating and orbiting and relative motion shifts the frequency, unless the transmitter compensates for that motion. A morphology with zero drift is therefore the most obviously terrestrial case in the set.

The detector rejects D2-1, but it cannot be rejecting it for being stationary, since it has no access to that reasoning. Two explanations remain: D2-1 is not tilted enough to match the signals it learned, or vertical lines are common interference in the training negatives and it learned to reject them. These were not separated here.

D2-1 and D2-5 are both terrestrial in origin, and both appear in all six panels. The detector rejects the first and accepts the second 63\% of the time at an amplitude of 0.5. The only difference between them is tilt, so the correct rejection of D2-1 shows nothing about the detector's ability to identify interference by its origin: it depended on morphology, whichever of the two explanations holds.

\subsection{Registered but under-weighted}
A missed signal can result from two different failures: the detector did not register it, or it registered it and gave it too little weight. The score alone cannot separate them.

Pairing each injection with its own clean background separates them (Section 4.1). Bright D1-1 lines did register: at an amplitude of 4.0 the score rose above the clean background's in 94\% of samples. But the rise shrank as the line brightened, so a brighter line raised the score less than a fainter one. For D1-4 the median change was close to zero, although its attention followed the injected path (on-path ratio 2.81$\times$).

The deletion test adds a local view. For D1-4 and D1-1, erasing the regions Grad-CAM ranked highest raised the score more than erasing random regions (10 of 12 samples each, p = 0.002 and p = 0.007), so those regions carried evidence against a detection. The detector therefore registered these signals and weighed parts of them against a detection; it did not push the score below the background. The evidence for D1-1 is weaker, as its on-path test was inconclusive.

This bears on what could fix it. If the detector had not registered the signal, more training examples of that morphology would be the obvious remedy. Since it did register it, the problem may also lie in how the detector weighs what it sees, which retraining could change but which was not tested here.

\subsection{Why this filter failed}
The filter failed for two reasons. First, its time averaging dilutes a drifting signal before ON and OFF are compared (Section 3.6), so a genuine signal shifts the statistic by only about 0.04. Second, the background already varies by more than that: the median ON-OFF asymmetry was 0.185 in clean backgrounds and 0.228 in flagged genuine signals.

ON and OFF panels are recorded minutes apart. During that interval the telescope repoints, the receiver can drift, and terrestrial interference varies in strength, which likely explains, though this was not tested, why the ON-OFF difference is substantial even when no genuine signal is present.

The veto threshold of 1.0266 sits in the upper tail of the background distribution, and only 1.4\% of genuine detections exceed it. Any cutoff high enough to reject interference therefore also rejects genuine signals.

This is not a matter of threshold selection. Among flagged detections, the statistic separates genuine signals from D2-5 interference with an AUC of only 0.60, so no threshold value separates them, hence no good choice exists.

The cadence information is present in the data, but this statistic does not recover it. A path-aware comparison, which integrates along the signal's drift before comparing ON and OFF as de-Doppler pipelines do, or a detector that learns what a genuine cadence looks like, may recover it; neither was tested here. This result therefore shows that a time-averaged ON-OFF statistic fails, not that post-hoc cadence checks in general cannot work.

\subsection{Operational consequences}
For anyone applying this detector, the dangerous case is faint interference, not bright. D2-5 produced a false positive rate of 0.627 at an amplitude of 0.5 but only 0.020 at an amplitude of 4.0.

The same amplitude range is where D1-1 and D1-2 are most detectable: they reached their highest recall, 0.410 and 0.533, at an amplitude of 0.5. For these morphologies, genuine detections and false alarms therefore fall into the same narrow amplitude band, and any operating point that captures genuine faint signals also admits the faint interference.

\section{Limitations}
The detector was trained once, with one random seed and one validation fold, so the failures reported here belong to that training run; whether they persist across retraining was not tested. The attention analysis rests on twelve samples per group, which is enough to show a consistent pattern but not to measure its size precisely. The on-path test for D1-5 used a null built from other D1-5 paths, which share one fixed curved form and overlap, so its ratio is biased towards 1.0; the true value may be higher, which makes the absence of attention to D1-5 the least certain result in this paper. The attention claim for D1-1 is also weaker than for D1-4: its on-path test was inconclusive, so it rests on the attention maps and the deletion test alone. All signals were synthetic, injected into real backgrounds, and covered ten morphologies; a real technosignature or a new kind of interference could look like none of them. Each Grad-CAM cell spans 16 channels, so attention on a line means attention in the band containing it. Finally, only one post-hoc filter was tested, and its time averaging made it insensitive to drifting signals by design.

The injections were simplified (Section 3.3): each ON panel repeats the same path, the 300 backgrounds were shared by every family and mostly seen in training, and they also served as the clean controls. The Direction 1 morphologies were assumed, not shown, to be absent from the training data. And the claim that the missed signals do not depend on the OFF panels, while the false alarms do, follows from the input design but was not tested against a detector that reads all six panels.

The fall in recall at high amplitude is also unexplained. A value-range artefact can be ruled out: the data reach values of about 37, with a 99.9th percentile near 12.6, so an injected line of amplitude 4 is well inside the normal range. Two hypotheses remain untested. The competition normalised each sample after injecting its signals, so a bright training signal would have reduced the apparent strength of the background around it, whereas my injections were not renormalised; bright injections may therefore be unlike anything the detector saw. Alternatively, genuine signals in the training data may have been fainter than the interference around them, so the detector learned brightness as a sign of interference. Renormalising each injected sample before scoring would test the first.

\section{Conclusion}
A CNN detector, trained once on Breakthrough Listen data and then frozen, was probed with 15,000 injections in two directions. It missed genuine signals whose shape it did not recognise, lost recall as signals brightened even for the kind of signal it was trained on, flagged one family of interference about as readily as it detects real signals, and attended to some signals it went on to reject. A pre-specified, time-averaged ON-OFF filter removed the false alarms only by removing the detections with them. The false alarms follow from one design choice: the detector reads only the ON panels, so it can judge how a signal looks but not where it comes from. The missed signals and the fall with brightness are separate problems, of training coverage and of an untested brightness effect. The evidence needed for that judgement is already collected in every cadence, and a detector that reads all six panels, together with a drift-aware cadence check, are the natural next steps.

\section*{Data and code availability}
The training and experiments notebooks, the injection manifest, the per-sample results, and the frozen model weights are available at \url{https://github.com/OmerImran/seti-two-directional-failure}.

\section*{Acknowledgements}
I used Claude (Anthropic) as an assistant during this project: to help debug code, revise a few portions of the text, and review the manuscript. The research question, experimental design, pre-specified decision rules, and all analysis decisions are my own. I ran all experiments myself on Kaggle, and every number reported comes from those saved outputs.

\appendix
\section{All detection rates}
\begin{table}[htbp]
\centering
\footnotesize
\setlength{\tabcolsep}{3pt}
\caption{Rate above the fixed threshold for every family and amplitude: recall for Direction 1, false positive rate for Direction 2. Each cell is computed on 300 injections; 95\% Wilson confidence intervals in brackets.}
\label{tab:allrates}
\begin{tabular}{lccccc}
\toprule
Family & 0.25 & 0.5 & 1.0 & 2.0 & 4.0 \\
\midrule
D1-1 extreme drift & 0.183 \scriptsize(0.144--0.231) & 0.410 \scriptsize(0.356--0.466) & 0.273 \scriptsize(0.226--0.326) & 0.077 \scriptsize(0.052--0.112) & 0.020 \scriptsize(0.009--0.043) \\
D1-2 quadratic & 0.453 \scriptsize(0.398--0.510) & 0.533 \scriptsize(0.477--0.589) & 0.340 \scriptsize(0.289--0.395) & 0.127 \scriptsize(0.094--0.169) & 0.017 \scriptsize(0.007--0.038) \\
D1-3 S-shaped & 0.253 \scriptsize(0.207--0.305) & 0.570 \scriptsize(0.513--0.625) & 0.750 \scriptsize(0.698--0.796) & 0.880 \scriptsize(0.838--0.912) & 0.900 \scriptsize(0.861--0.929) \\
D1-4 pulsed & 0.023 \scriptsize(0.011--0.047) & 0.017 \scriptsize(0.007--0.038) & 0.017 \scriptsize(0.007--0.038) & 0.003 \scriptsize(0.001--0.019) & 0.010 \scriptsize(0.003--0.029) \\
D1-5 dispersed burst & 0.013 \scriptsize(0.005--0.034) & 0.017 \scriptsize(0.007--0.038) & 0.010 \scriptsize(0.003--0.029) & 0.010 \scriptsize(0.003--0.029) & 0.010 \scriptsize(0.003--0.029) \\
\midrule
D2-1 zero drift & 0.013 \scriptsize(0.005--0.034) & 0.017 \scriptsize(0.007--0.038) & 0.010 \scriptsize(0.003--0.029) & 0.013 \scriptsize(0.005--0.034) & 0.010 \scriptsize(0.003--0.029) \\
D2-2 comb & 0.013 \scriptsize(0.005--0.034) & 0.017 \scriptsize(0.007--0.038) & 0.003 \scriptsize(0.001--0.019) & 0.007 \scriptsize(0.002--0.024) & 0.010 \scriptsize(0.003--0.029) \\
D2-3 intermittent & 0.010 \scriptsize(0.003--0.029) & 0.010 \scriptsize(0.003--0.029) & 0.010 \scriptsize(0.003--0.029) & 0.007 \scriptsize(0.002--0.024) & 0.013 \scriptsize(0.005--0.034) \\
D2-4 broadband burst & 0.013 \scriptsize(0.005--0.034) & 0.013 \scriptsize(0.005--0.034) & 0.013 \scriptsize(0.005--0.034) & 0.013 \scriptsize(0.005--0.034) & 0.010 \scriptsize(0.003--0.029) \\
D2-5 high drift & 0.460 \scriptsize(0.404--0.517) & 0.627 \scriptsize(0.571--0.679) & 0.307 \scriptsize(0.257--0.361) & 0.033 \scriptsize(0.018--0.060) & 0.020 \scriptsize(0.009--0.043) \\
\bottomrule
\end{tabular}
\end{table}

\end{document}